\documentclass[doublecol, figures]{epl2}

\title{Roughness correction to the Casimir force beyond perturbation
theory}

\author{Wijnand Broer\inst{1} \and George Palasantzas\inst{1} \and Jasper Knoester\inst{1}, \and
Vitaly B. Svetovoy\inst{2}}
\shortauthor{Wijnand Broer \etal}
\institute{\inst{1} Zernike Institute for Advanced Materials, University of Groningen, Nijenborgh 4, 9747 AG Groningen, The Netherlands\\
\inst{2} MESA$^+$ Institute for Nanotechnology, University of
Twente, P.O. Box 217, 7500 AE Enschede, The Netherlands}


\abstract{Up to now there has been no reliable method to calculate
the Casimir force when surface roughness becomes comparable with the
separation between bodies. Statistical analysis of rough Au
films demonstrates rare peaks with heights considerably larger
than the root-mean-square (rms) roughness. These peaks define the
minimal distance between rough surfaces and can be described with
extreme value statistics. We show that the contributions of high
peaks to the force can be calculated independently of each other
while the contribution of normal roughness can be evaluated
perturbatively beyond the proximity force approximation. The
developed method allows a reliable force estimation for short
separations. Our model explains the strong  hitherto unexplained deviation from the
normal Casimir scaling observed experimentally at short separations.
}

\begin{document}
\maketitle

\section{Introduction}
The Casimir force \cite{Cas48} attracts increasing attention
nowadays since modern technology allows dimension control at
distances $\leq 100$ nm where this force becomes operative (see
\cite{Cap07,Rod11} for a review). Indeed, modern
micro/nano-electromechanical (MEM/NEM) engineering is now being
conducted at the micron to nanometer scale and has attracted
interest in the Casimir force \cite{Bal07}. MEM devices such as
vibration sensors and switches are now routinely made with parts a
few micrometers in size, and have the right size for the Casimir
force to play a role. This is because MEM systems have surface
areas large enough but gaps small enough, for the force to draw
components together and possibly lock them permanently - an effect
known as stiction. Such permanent adhesion (in addition to
capillary adhesion due to the water layer) is a common cause of
malfunctioning of MEM devices \cite{Ser98,Buk01a,Buk01}.

In this range of separations the force appears mainly due to
quantum fluctuations of the electromagnetic field (zero-point
field) in the interacting bodies while at larger distances
classical (thermal) fluctuations become increasingly important
\cite{DLP,LP9}. The famous Casimir formula $F_C=(\pi^2/240)(\hbar
c/d^4)$ gives the force (per unit area) at temperature $T=0$
between two ideally reflecting semi-spaces separated by the
distance $d$. The force measured in recent experiments (see
\cite{Rod11} for a review) can deviate significantly from the
ideal case because the temperature is finite, the bodies are not
ideal reflectors, and the distance between them is not well
defined. Considerable efforts were made to improve the Casimir
formula. Indeed, the more detailed description is based on the
Lifshitz formula, which accounts for actual optical properties of
interacting bodies and nonzero temperature. The optical data were
included in the calculational procedure \cite{Lam00,Sve08}.
Although the thermal correction to the Casimir force is rather
controversial \cite{Mil04,Sus11}, it is not important for the short
distances discussed in this paper.

An important correction to the Casimir
force that is not accounted for by
the Lifshitz formula is the roughness correction. The
surfaces of real bodies are rough, which makes the distance
between them not well defined. The first attempts to account for
roughness \cite{Kli96} were based on the proximity force
approximation (PFA). In this approximation the real surfaces are
replaced by flat patches and the force was calculated as the sum of forces between opposite patches, treating such pairs as parallel plates. For the
dispersive forces the PFA was applied for the first time by
Derjaguin \cite{Der34,Der56}. The approximation is justified when
the separation $d$ is much smaller than the local curvature radius
and size of patches. It is well suited for smooth large bodies,
but works worse for roughness corrections. It was noted
\cite{Gen03} that in order to apply the PFA to rough bodies the
roughness correlation length $\xi$ (typical features size on the
surface) must be larger than the separation, $\xi\gg d$. Then the
result found in \cite{Kli96} will be true for small  root mean
square (rms) roughness, $w\ll d$. However, in most of the
experimental situations the condition $\xi\gg d$ is broken and
more elaborate theory has to be used to calculate the roughness
correction. This theory was developed in refs.
\cite{Net05a,Net05b}. It treats the roughness contribution through
second order perturbation theory in $w/d$. The theory
showed larger corrections than those predicted within the PFA. In
fact, the correction is very important at short separations and
has to be carefully included for interpretation of the force
experiments exploring short distance ranges.

The Casimir forces between a gold covered sphere and plates of
different roughness were measured for separations from 20 to 200
nm \cite{Zwo08}. The films with larger rms roughness at short
separations demonstrate significant (more than 100\%) deviations
from the theoretical expectations based on the perturbative
roughness correction. Empirically it was established that the
minimal distance between two rough bodies (distance upon contact)
is $d_0\approx 3.7(w+w_{sph})$, where $w_{sph}$ is the sphere's rms roughness. Because $d_0$ is the minimum separation distance, the perturbative
correction must be smaller than $K (w+w_{sph})^2/d_0^2=0.07K$,
where $K\sim 10$ is a large numerical factor (due to sharp behavior of
the force with the distance). It was concluded \cite{Zwo08} that
at short separations the perturbation theory fails. These
experimental results still did not get a theoretical explanation.
Moreover, we are facing a problem: there is no reliable
method to estimate the roughness correction when $d$ becomes
comparable with the rms roughness $w$. In this paper we propose a
method to address this problem by combining the PFA and
perturbation theory approaches. Although we will prove the
applicability of this method specifically for gold films, we
believe that similar approaches can be developed for other
materials, after detailed analysis of the roughness statistics
obtained, e.g., in terms of scanning probe microscopy techniques.

\section{Statistics of rough surfaces}

The distance upon contact $d_0$ was discussed in detail for gold
films \cite{Zwo09}. The films deposited with different thicknesses
have different rms roughnesses due to kinetic roughening
processes. For all these films atomic force microscope (AFM)
images were recorded for large area (of up to $40\times 40$
$\mu\textrm{m}^2$) with lateral resolution of 4-10 nm. This
information allows a detailed analysis of the roughness
statistics. The probability to find  a height of a local feature
smaller than some value $z$ can be presented in a general form
\begin{equation}\label{distr_P}
    P(z)=1-e^{-\phi(z)},
\end{equation}
where for convenience we introduced the "phase" $\phi(z)$ as
nonnegative and nondecreasing function of $z$. The phase describes the roughness distribution in a convenient way, which makes it possible to calculate the contributions from peaks and troughs as will be shown later. It was already noted \cite{Zwo09} that the cumulative distribution $P(z)$ for
gold films cannot be described satisfactorily by any known
distribution at all $z$ but asymptotically at large $z$ it can be
fitted with generalized extreme value distributions
\cite{Col01}.

\begin{figure}[ptb]
\begin{center}
\onefigure[width=0.49\textwidth]{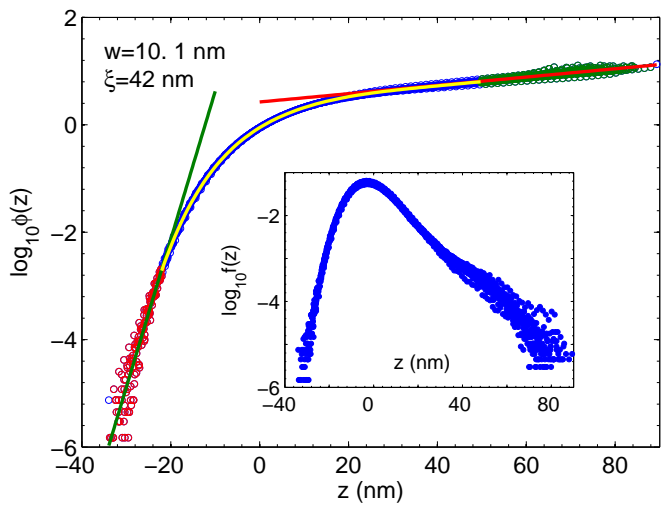}
\vspace{-0.3cm} \caption{(Color online) The ``phase'' as a function of $z$ for a 1600 nm gold film. The open circles are the actual data
extracted from the AFM image using eq. (\ref{phi_def}). At large
positive and large negative heights $\log_{10}\phi(z)$ is well
fitted by linear functions of $z$ as is shown by the straight
lines. The curved line is a polynomial fit at intermediate $z$.
The inset shows the probability density function $f(z)$. It
demonstrates significant deviation from a normal distribution. }
\label{fig1} \vspace{-0.9cm}
\end{center}
\end{figure}

We performed a special analysis of the AFM surface data presented in ref. \cite{Zwo09} to reveal the best asymptotic distribution at large $|z|$. In this limit the phase $\phi(z)$ is much more convenient for analysis than $P(z)$. This is because $P(z)$ approaches very fast 0 or 1 in the limit
$|z|\rightarrow\infty$. Indeed, we can present the phase as
\begin{equation}\label{phi_def}
    \phi(z)=-\ln\left[1-P(z)\right],
\end{equation}
where $P(z)$ is extracted directly from the images.  The function
$\phi(z)$ for an 1600 nm thick gold film is shown in fig.
\ref{fig1}. The inset shows the probability density function
$f(z)=dP/dz=(1-P)d\phi/dz$. Similar behavior is realized for all
investigated gold films. It is clear that for large positive $z$
the logarithm of the phase can be fitted with a linear function
\begin{equation}\label{ln_phi}
    \ln\phi(z)=A+Bz,\ \ \ z\rightarrow\infty
\end{equation}
and similarly for large negative $z$. With this $\phi$ the
probability to find a feature larger than $z$ behaves asymptotically as a double exponential
\begin{equation}\label{Gum}
    1-P(z)\sim\exp\left[-\exp\left(\frac{z-\mu}{\beta}\right)\right],
\end{equation}
where $\beta$ and $\mu$ are the scale and location parameters
respectively. This behavior is a characteristic feature of the
Gumbel distribution \cite{Gum04}, which is an example of extreme
value statistics. In this paper only gold films were analyzed and
therefore we cannot draw conclusions on the roughness statistics
of other materials. However, the extreme character of the
statistics allows us to hope that this behavior is more general.

\section{Roughness correction to the Casimir force}

We can imagine a rough surface as a large number of asperities
with heights $\sim w$ and lateral size $\xi$, and occasional high
peaks and deep troughs. These peaks (or troughs) are high in the
sense that their height is considerably larger than $w$, say
$>3w$. The situation can be visualized as a lawn covered with
grass and occasional high trees standing here and there. In this
paper we propose a method to calculate the roughness correction to
the Casimir force on the basis of this separation. Namely, the
asperities with the height $\sim w$ can be taken into account
using perturbation theory, without the use of the proximity
force approximation. On the other hand, for high peaks the local
distance between interacting bodies becomes considerably smaller
and one cannot use perturbation theory anymore. Because high peaks are rare the average distance $l$ between them is
large. If this distance is so large that $l\gg d$, then we can
calculate the contribution of these peaks independently of each
other, as it is assumed in  the PFA. However, this contribution has
to be calculated beyond the perturbation theory. It has to be
stressed that the interaction of a separate peak with a flat surface
can be taken into account precisely using developed numerical or
analytical methods \cite{Rod11}.

The number of asperities $N$ with the height $d_1>3w$ and lateral
size $\xi$ on the area $L^2$ is given by the equation \cite{Zwo09}
\begin{equation}\label{N_def}
    N=\frac{L^2}{\xi^2}e^{-\phi(d_1)}.
\end{equation}
The average distance between these peaks is
\begin{equation}\label{l_def}
    l=\frac{L}{\sqrt{N}}=\xi e^{\phi(d_1)/2}.
\end{equation}
In order to fulfill the condition of PFA applicability $l\gg d$, we can choose the parameter $d_1$ from the interval $3w<d_1<d_0$,
where $d_0$ is the maximal peak on the area $L^2$. The best
choice for $d_1$ will be discussed later. Similarly, one can
introduce the average distance $l^{\prime}$ between deep troughs
\begin{equation}\label{l'_def}
    l^{\prime}=\frac{L}{\sqrt{N^{\prime}}}=\frac{\xi}
    {\sqrt{\phi(-d_1^{\prime})}},
\end{equation}
where $d_1^{\prime}$ has to be chosen from the interval
$3w<d_1^{\prime}<d_0^{\prime}$ to fulfill the condition
$l^{\prime}\gg d$ and $d_0^{\prime}$ is the deepest trough on the
area $L^2$.

Here we consider the general case where we are interested in the Casimir force between two plates with rough surfaces. As was explained in ref.
\cite{Zwo09} this is equivalent to the interaction of a smooth plate
with a rough one, which has the combined roughness topography
$h(x,y)=h_1(x,y)+h_2(x,y)$, where $h_{1,2}(x,y)$ are the
topographies of the interacting plates 1 or 2. Therefore all the
equations above have to be applied to the combined roughness profile
$h(x,y)$.

Let us assume for a moment that the PFA can be applied to any
roughness topography. Then the force between the plates can be
calculated using the standard definition of the averaged function
\begin{equation}\label{aver}
    {\cal F}(d)=\int\limits_{d_1}^{d_0}\ldots
    +\int\limits_{-d_0^{\prime}}^{-d_1^{\prime}}\ldots +
    \int\limits_{-d_1^{\prime}}^{d_1}dz
    f(z)F(d-z),
\end{equation}
where $f(z)$ is the probability density function, and we separated
high peaks (first integral), deep troughs (second integral), and the
normal roughness contribution (third integral). For the moment we
do not specify the force between the interacting patches $F(d)$
separated by the distance $d$. The last term can be calculated
using the perturbation expansion
$F(d-z)=F(d)-F^{\prime}(d)z+F^{\prime\prime}(d)z^2/2!+\ldots$ and
we find for this term
\begin{equation}\label{PExp}
    \int\limits_{-d_1^{\prime}}^{d_1}\ldots=
    F(d)\int\limits_{-d_1^{\prime}}^{d_1}
    dzf(z)+\frac{F^{\prime\prime}(d)}{2!}
    \int\limits_{-d_1^{\prime}}^{d_1}dzf(z)z^2.
\end{equation}
The first and second integral on the right are 1 and $w^2$,
respectively, if one extends the integration limits to infinity.
When the applicability of the PFA breaks down the second term in
(\ref{PExp}) (with infinite limits) can be generalized as follows
\cite{Gen03}
\begin{equation}\label{PT}
    {\cal F}_{PT}(d)=\frac{F^{\prime\prime}(d)}{2!}
    \int\frac{d^2k}{(2\pi)^2}\rho(kd)\sigma(k).
\end{equation}
Here $\sigma(k)$ is the Fourier spectrum of the roughness
correlation function. The function $\rho(kd)$ measures the deviation
from the PFA. When the PFA is applicable this function is $\rho(kd)=1$ and we reproduce eq. (\ref{PExp}). Outside of the PFA applicability we can use for $\rho(kd)$ expressions found in \cite{Net05a,Net05b}.

The term ${\cal F}_{PT}(d)$  is the roughness contribution to the
force treated as a perturbation theory correction. As we know
already the contribution of high peaks (or deep troughs) may not
be accounted for by the perturbation theory. However, in this case we
can account for the peaks (troughs) independently and the contribution
can be presented as the first (second) term in (\ref{aver}).
Taking into account the change of the integration limits we find
the contributions due to high peaks, ${\cal F}_{PFA}(d)$, and due
to deep troughs, ${\cal F}_{PFA}^{\prime}(d)$,

\begin{eqnarray}
  \label{PFA}\lefteqn{{\cal F}_{PFA}(d) =}\nonumber\\ &\int\limits_{d_1}^{d_0}dz f(z)\left[F(d-z)-F(d)+F^{\prime}(d)z-\frac{F^{\prime\prime}(d)}{2!}z^2\right], \\
  \label{PFA'}\lefteqn{{\cal F}^{\prime}_{PFA}(d)=}\nonumber\\ &\int\limits_{-d_0^{\prime}}^{-d_1^{\prime}}dz f(z)
    \left[F(d-z)-F(d)+F^{\prime}(d)z-\frac{F^{\prime\prime}(d)}{2!}z^2\right].
\end{eqnarray}
The final expression for the force that includes the total
roughness contribution can be presented as
\begin{equation}\label{force}
    {\cal F}(d)=F(d)+{\cal F}_{PT}(d)+{\cal F}_{PFA}(d)+
    {\cal F}^{\prime}_{PFA}(d).
\end{equation}
Here $F(d)$ is the force between flat surfaces and the other three
terms are the different roughness corrections.

The same force $F(d)$ is used to calculate ${\cal F}_{PFA}(d)$ and ${\cal
F}^{\prime}_{PFA}(d)$, which implies that high peaks are
described as pillars with flat faces. However, this approximation
is not necessary. If the peaks can be considered as independent,
then the interaction of each peak with the flat surface can be
described precisely (numerically) or approximately with an
appropriate force $\tilde{F}(d)$ in eqs. (\ref{PFA}) and (\ref{PFA'}),
taking into account the actual geometry of the peak. For example,
high peaks can be considered as pillars with spherical caps of
radius $\xi/2$. As we will see below for the description of the
experiment \cite{Zwo08} it is sufficient to use the simplest model
for the peaks (flat faces).


At this point an important question is: with what precision can we calculate the roughness corrections? The term $-F^{\prime\prime\prime}(d)z^3/3!$, which is neglected in the
Taylor expansion of $F(d-z)$, allows an estimation of the error in
${\cal F}_{PT}$. In the distance range that we are interested in
here, $20<d<100$ nm, the force $F(d)$ behaves with the distance as
$F(d)=A/d^{\alpha}$, where $A$ is a constant and $\alpha\approx
3.5$ \cite{Pal10}. Then the error is estimated as
\begin{equation}\label{err_PT}
    \Delta{\cal F}_{PT}=\gamma\frac{\alpha(\alpha+1)(\alpha+2)}{3!}
    \left(\frac{w}{d}\right)^3F(d),
\end{equation}
where $\gamma$ is the skewness of the distribution $f(z)$. The
data shown in fig. \ref{fig1} give the largest $\gamma=1.285$
among the investigated films and we estimate the error as
$\Delta{\cal F}_{PT}\approx 18.55(w/d)^3F(d)$. The latter means
that the perturbation theory correction has meaning at least for
$d/4>w$. The minimal distance between rough surfaces $d_0$ depends
on the area of nominal contact $L^2$, but even for $L$ as small as
$1\ \mu \textrm{m}^2$ the condition $d_0/w>4$ is usually fulfilled
\cite{Zwo09}. Therefore, we can now draw the important conclusion
that the perturbation theory correction (\ref{PT}) has meaning
up to the point of contact between interacting rough surfaces.

The precision with which we calculate the contribution of the high
peaks is defined by the condition of applicability of the PFA to
these peaks. This condition  is $l(d_1)\gg d$ and we have for the
error
\begin{equation}\label{err_PFA}
    \Delta{\cal F}_{PFA}=\left(d/l\right){\cal F}_{PFA}.
\end{equation}
As we already mentioned we have to choose $d_1$ such that the
condition $l(d_1)\gg d$ is true and, therefore, the correction
(\ref{PFA}) makes sense. Similarly, we can define the error
$\Delta{\cal F}_{PFA}^{\prime}$ for the contribution of deep
troughs (\ref{PFA'}). The relative error in (\ref{PFA}) increases
with the distance, but we have to keep in mind that ${\cal
F}_{PFA}$ decreases very fast with $d$ and the absolute error
stays small. The parameters $d_1$ and $d_1^{\prime}$ can be chosen
rather arbitrarily if the conditions $l(d_1)\gg d$ and
$l(d_1^{\prime})\gg d$ are fulfilled. A practical recipe could be
$d_1=\max\left[3w,(d_0+w)/2\right]$ and
$d_1^{\prime}=\max\left[3w,(d_0^{\prime}+w)/2\right]$. It has to
be noted that the contribution of deep troughs is always small, but
we keep it for the sake of generality.

\section{Results}

The roughness corrections (\ref{PT})-(\ref{PFA'}) were deduced for
the force between two rough parallel plates. In most of the
experimental configurations the sphere-plate geometry is used. We
can find the result for this configuration if the sphere's radius
is large, $R\gg d$.  This condition is typically true when the
roughness effect is appreciable and we can apply the PFA to the
total force ${\cal F}(d)$. The same equations
(\ref{PT})-(\ref{force}) can be applied but now we have to
understand $F(d)$ as the force between a smooth sphere and a
smooth plate, approximated by
\begin{equation}\label{PFAcurv}
F(d)=2\pi{}RE(d)\qquad{}R\gg{}d,
\end{equation}
where $E(d)$ is the Casimir-Lifshitz energy per unit area for the
parallel plate configuration \cite{LP9}. We neglect the thermal
effect ($T=0$) at short separations \cite{Mil04}. However, we use
measured optical properties of gold films \cite{Sve08} to account
for the actual material properties.

We evaluated the force and all the roughness corrections to
compare it with the experimental data \cite{Zwo08}. The Lifshitz
force $F(d)$ was calculated for the sphere radius $R=50\ \mu
\textrm{m}$ using the optical data for sample 3 in \cite{Sve08}.
The roughness effect was estimated for 800, 1200, and 1600 nm Au
films and a Au covered sphere. Here we present the results for the 1600
nm film. The roughness characteristics for combined sphere-plate
AFM images were presented in ref. \cite{Zwo09}. For the 1600 nm
film they are: rms roughness $w=10.1$ nm, the correlation length
$\xi=42$ nm, and the distance upon contact $d_0=50.8\pm 1.3$ nm.
The last value was determined by electrostatic calibration
\cite{Zwo08}. It is preferable to use this value, because $d_0$
determined from the roughness topography has a larger uncertainty
\cite{Zwo09}. We used $d_1=(w+d_0)/2=30.5$ nm. According to eq.
(\ref{l_def}) it corresponds to the average distance between high
peaks, $l\approx 380$ nm. Note that the effective area of
interaction is $L^2$, with $L=2100$ nm \cite{Zwo09}. For
deep troughs the calculation details are less important. For the
given $L$ we found $d_0^{\prime}=24.6$ nm. Since $d_0^{\prime}<3w$
the troughs are not deep enough and can be taken into account
perturbatively. Therefore, in this specific case there is no need to introduce ${\cal F}^{\prime}_{PFA}$.

\begin{figure}[ptb]
\begin{center}
\onefigure[width=0.49\textwidth]{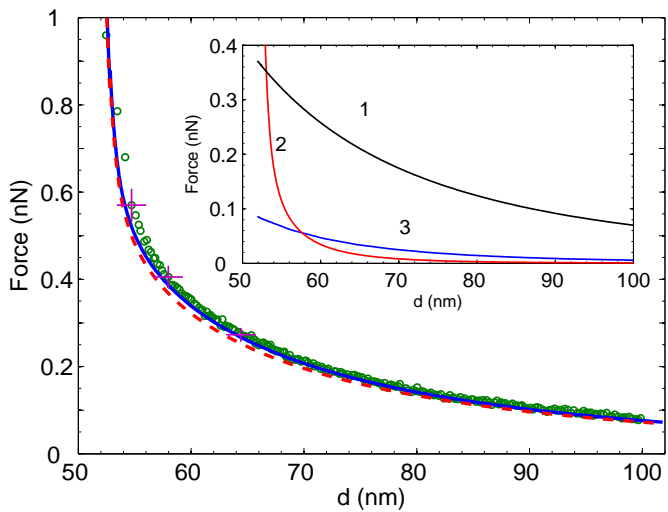}
\vspace{-0.3cm} \caption{(Color online) The force between a Au
covered sphere ($R=50$ $\mu \textrm{m}$) and a plate (1600 nm thick
Au). The open (green) circles are the experimental data from
\cite{Zwo08}. The vertical and horizontal bars show the
experimental errors for a few points. The solid (blue) line is the
result of our model. Naive application of the PFA to the force between rough bodies based on eqs. (\ref{aver}) and (\ref{PFAcurv}) is shown by the dashed
(red) curve. The inset shows different components of the force.  1
(black) is the force $F(d)$ between smooth surfaces calculated
according to the Lifshitz formula. 2 (red) is the contribution of
the high peaks according to eq. (\ref{PFA}). 3 (blue) is the
perturbation theory correction according to eq. (\ref{PT}). The
sum of all three curves gives the solid line in the main panel. }
\label{fig2} \vspace{-0.9cm}
\end{center}
\end{figure}

The results are presented in fig. {\ref{fig2}}. One can see that
the solid (blue) line, which shows the result of our approach, is
in agreement with the experimental data within the experimental
errors. This is in contrast with the perturbation theory approach
that failed to explain the data \cite{Zwo08}. This is demonstrated in
the inset, which shows different components of this force. At
short distances the contribution of high peaks (2, red) is so
large that it dominates the whole force. In this case a few peaks
become very close to the opposite body so that the force diverges
at $d\rightarrow d_0$. There can be very few high peaks but their
contribution cannot be neglected. On the other hand the
contribution of high peaks disappears very fast when the distance
becomes larger.

We used two different models to calculate the contribution of high
peaks in eq.(\ref{PFA}). In the first model the peak was considered as a
pillar with a flat face. In the second model the peak had a
spherical cap of radius $\xi/2$. The interaction of the cap with a
plate was taken into account according to ref. \cite{Dur09}, where the
proximity force approximation is not used. We found a negligible
difference between the two models of peaks. The reason is the
following: When the distance $d-d_0 \sim \xi$, then the 
contribution of the peaks is very small due to their small area of interaction. When
$d$ approaches $d_0$ or $d-d_0\ll \xi$, the contribution of high
peaks becomes significant, but the shape of the peaks is not
important anymore, because the PFA is valid in this limit.

Naive application of the proximity force approximation according
to eqs. (\ref{aver}) and (\ref{PFAcurv}) gives the dashed line (red) in fig.
\ref{fig2}. It is interesting to note that this line is also in
agreement with the experimental data. At the shortest separations
both curves coincide, because the dominating high peaks can be
treated with the PFA. At larger distances the perturbative
contribution becomes important and the PFA result lies below the
solid line as it should be \cite{Gen03}. However, the difference
between these two curves is within the experimental errors. Perturbation theory accounts for the non-additivity of the Casimir force, whereas the PFA assumes it is additive. So this difference provides an indication of the \emph{effect} of the non-additivity in the roughness correction. It can be concluded that within the experimental error the experiment in ref. \cite{Zwo08} was not sensitive to this non-additivity.

\section{Conclusions}
In conclusion, we developed a reliable method to include the
effect of roughness of interacting bodies in the Casimir force at
short distances when perturbation theory fails. It was established
that roughness of gold films can be described asymptotically (for
high peaks or deep troughs)  by extreme value statistics. In this
case the rough surface can be presented as a large number of
asperities with heights of the order of the rms roughness and a
few occasional peaks, which are much higher than the rms
roughness. The distance between high peaks is large so that one
can calculate their contribution for each peak separately (using
the PFA). The smaller asperities can be calculated using
perturbation theory beyond the PFA. The contribution of high peaks
is extremely important for short separations, where it dominates
not only the perturbative roughness correction but also the force
as a whole. Therefore, our result is interesting not only for
the Casimir force but also for the problem of adhesion between
surfaces in general \cite{Del05}, including wet environments
\cite{Del07,Zwo08b,Per08}.
\par We repeat that the method presented here solves the significant discrepancy between measurements of the Casimir force at short separations, and the results of perturbation theory \cite{Zwo08}.

\acknowledgments The authors benefited from exchange of ideas
within the ESF Research Network CASIMIR. We would also like  to
thank P. J. van Zwol and B. J. Hoenders for useful discussions.


\end{document}